\documentclass[12pt]{article}
\usepackage{amsmath}
\usepackage{amssymb}

\newcommand{\A}{\mathbf{A}}
\newcommand{\B}{\mathbf{B}}
\newcommand{\biq}{\mathop{\pmb{Q}}}

\newcommand{\C}{\mathbf{C}}

\newcommand{\D}{\mathbf{D}}
\newcommand{\bF}{\mathbf{F}}
\newcommand{\F}{\mathbf{F}}
\newcommand{\Fop}{\mathbf{F}\mathrm{(op)}}
\newcommand{\Fv}{\mathbf{F}(v)}
\newcommand{\I}{\mathbf{I}}

\newcommand{\boldg}{\mathbf{g}}
\newcommand{\bfsigma}{\boldsymbol{\sigma}}
\newcommand{\X}{\mathbf{X}}
\newcommand{\bQ}{\mathbf{Q}}
\newcommand{\btau}{\boldsymbol{\tau}}
\newcommand{\bU}{\mathbf{U}}

\newcommand{\bV}{\mathbf{V}}
\newcommand{\Q}{\mathbf{Q}}
\newcommand{\U}{\mathbf{U}}
\newcommand{\V}{\mathbf{V}}

\newcommand\diag{\mathrm{diag}}
\newcommand{\M}{\mathbf{M}}
\newcommand{\N}{\mathbf{N}}

\begin{document}

\title{Electric Charge as a Vector Quantity}
\author{Gerald L. Fitzpatrick}
\date{PRI Research and Development Corp.\\
12517 131 Ct.\  N.\ E.\\
Kirkland, WA 98034\\
425--820--1905\\
glfitzpatrick@yahoo.com}
\maketitle

\begin{abstract}

\noindent Starting with the premise that the electric charge associated with fundamental fermions (quarks and leptons) can, under certain circumstances, be appropriately represented as a real \emph{internal} 
2-vector, the mathematical ``machinery'' implicit in the associated internal 
2-space is shown to apply to \emph{all} fundamental fermions. In particular, it is shown that \emph{flavor eigenstates}, \emph{flavor doublets} and \emph{families} of fundamental fermions can all be represented in the 2-space, and that such things as internal \emph{colors}, \emph{family replication}, and the observed \emph{number} (three) of families, are more-or-less implicit in the new 2-space description. 
Moreover, the model predicts that, unlike the case in the standard model, particles such as the $u$, $c$ and $t$ quarks are characterized by significant internal (topological and other) differences. Similar differences may help explain recent observations of (nearly) maximal $\nu_{\mu}-\nu_{\tau}$ mixing.
\end{abstract}

\setcounter{section}{0}
\setcounter{subsection}{-1}

\section{Introduction} 

If quantum effects are ignored it is completely appropriate to treat electric charge in \emph{external} spacetime as a \emph{scalar} quantity. In particular, in  4-dimensional spacetime, the electric charge carried by an isolated particle is appropriately treated as a Lorentz invariant 4-scalar \cite{1}. However, when quantum effects are taken into account it is not at all clear that the \emph{internal}
description of electric charge should be limited to scalar quantities. The purpose of this paper will be to argue that in the case of fundamental fermions (quarks and leptons) there are good reasons for treating electric charge as a \emph{vector} quantity associated with a new (abstract) \emph{internal} 2-space \cite{2}. The motivation for taking such an unusual step is that it will be shown to lead to an \emph{extension} of the standard model 
description of quarks and leptons. 

Before embarking on this exploration, it is appropriate to briefly review the conventional description of flavor doublets of fundamental fermions. Certain aspects of the conventional description via the $SU(2)$ isospin formalism will suggest the new description.

\subsection{The conventional description of flavor doublets}

According to the standard model of particle physics, all left-handed quarks and leptons (right-handed antiquarks and antileptons) are members of $SU(2)$ weak isospin \emph{doublets} [3].
To be specific let us limit the present discussion to the first-family quark states $|u\rangle$ and $|d\rangle$, which constitute an $SU(2)$ weak isospin doublet (also called a flavor doublet). These states are properly to be thought of as being two \emph{different} (isospin) states of a \emph{single} quark field [4]. 

Using the conventional isospin language, there exists an isospin (vector) \emph{operator}
$\btau=\frac{1}{2}\bfsigma$, where $\bfsigma=(\bfsigma_1, \bfsigma_2, \bfsigma_3)$ is a vector form whose components are the familiar Pauli matrices $\bfsigma_1, \bfsigma_2$ and $\bfsigma_3$.
 All physically observable states carrying (weak) isospin (e.g., $|u\rangle$ and $|d\rangle$) must be simultaneous eigenstates of the square of the total isospin vector $\btau^2=\btau^2_1+\btau^2_2+\btau^2_3$, and the third-component of isospin $\tau_3$, i.e., they  take the form $|\btau^2, \tau_3\rangle$. The states $|u\rangle$ and $|d\rangle$ are said to span a two-dimensional Hilbert space, and to constitute a 2-dimensional representation $\D^{(1/2)}$ of $SU(2)$.
In particular, given that the states $|u\rangle$ and $|d\rangle$ are eigenstates of \begin{equation}\label{eqn1}
\btau_3=\frac{1}{2}\bfsigma_3,
\end{equation}
where
\begin{equation}\label{eqn2}
\bfsigma_3 = \left(\begin{array}{cc}
1 & 0 \\
0 & -1 \end{array}\right),
\end{equation}
we have the  eigenvalue equations (use $\btau_3|\btau^2,\tau_3\rangle = \tau_3|\btau^2,\tau_3\rangle$ or $\btau_3|\frac{3}{4}, \pm\frac{1}{2}\rangle = \pm\frac{1}{2}|\frac{3}{4}, \pm\frac{1}{2}\rangle$ and/or the column-vector forms $|u\rangle=\{1,0\}$ and $|d\rangle=\{0,1\}$)
\begin{equation}\label{eqn3}
\left.\begin{array}{rcl}
\btau_3|u\rangle & = & +\frac{1}{2}|u
\rangle \\[3mm]
\btau_3|d\rangle & = & -\frac{1}{2}|d\rangle \end{array}\right\}.
\end{equation}

Now the \emph{electric charge} of the quark field in either of the states $|u\rangle$ or $|d\rangle$ is given by operating on these states with the operator for electric charge, which may be expressed in this particular case as
\begin{equation}\label{eqn6}
\biq=\btau_3+\frac{1}{6}.
\end{equation}
Then we have for the electric charges of these \emph{two} possible states of the quark field
\begin{align}
\biq|u\rangle & = q_1\;|u\rangle \label{eqn7} \\
\intertext{and}
\biq|d\rangle & = q_2\;|d\rangle, \label{eqn8}
\end{align}
where
\begin{equation}\label{eqn9}
q_1=+\frac{2}{3}\hbox{ and }q_2=-\frac{1}{3}
\end{equation}
are the specific electric charges in question.

While the foregoing description is certainly correct as far as it goes it is, nevertheless, unnecessarily restrictive. In particular, we will show in the next section that the foregoing properties of flavor doublets are suggestive of a (complementary) extension of this conventional description, which treats electric charge as a real \emph{vector} quantity in a new \emph{internal} 2-dimensional linear vector space.

\subsection{Electric charge as a vector quantity}

It is implicit in the description of $SU(2)$ 
isospin
\emph{doublets} that 
if one
 knows the state $|u\rangle$ one can (must) infer the existence of a second state $|d\rangle$, and vice versa. The $2\times 2$ matrix form of the charge operator $\biq$, and the isospin operator $\btau_3$  makes this abundantly clear. 

If one has a state with $\tau_3=+\frac{1}{2}$, and one knows that one is dealing with an $SU(2)$ isospin
doublet field, then one must also have a state with $\tau_3=-\frac{1}{2}$. As a consequence of these simple facts, one can have \emph{simultaneous} knowledge of the two \emph{electric charges} $q_1$ and $q_2$ associated with the two states $|u\rangle$ and $|d\rangle$, respectively, given \emph{either} one of the states $|u\rangle$ or $|d\rangle$. 

In a certain sense then, there could be a physical meaning to a geometric object defined by the ordered pair of observable (real numbers) \emph{electric charges} associated with the states $|u\rangle$ \emph{and} $|d\rangle$, namely, a \emph{real} 2-vector or ``charge vector''
\begin{equation}\label{eqn10}
\bQ=\{q_1, q_2\}.
\end{equation}
However, it should be stressed that if the 2-vector $\Q$ were to be ``carried'' by, or ``associated'' with, each of the states $|u\rangle$ \emph{and} $|d\rangle$ there would be definite nontrivial physical consequences. 

Clearly, the assignment of $\Q$ to $|u\rangle$ \emph{and} $|d\rangle$ would mean that these states carry \emph{additional} information (besides isospin) in the quantum sense [5]. In particular, because the 2-vector $\bQ$ has to ``live'' in some abstract \emph{internal} 2-space, all of the mathematical ``machinery'' associated with this 2-space would have to be taken into account (e.g., the 2-space metric and various internal transformations in the 2-space) when describing fundamental fermions. In short, this new description would promise a far \emph{richer} internal structure than that implied by the description of flavor doublets using $SU(2)$ alone. 
This implied richness constitutes nothing less than a possible \emph{extension} of the standard model [3, 6], and encourages us to seriously consider the idea of representing electric charge (of fundamental fermions) by an \emph{internal} 2-vector. 

In the next section we will show how this idea leads naturally to, among other things, an explanation for flavor doublets in other families, i.e., to an explanation for \emph{family replication}.

\section{Consequences of Treating Electric Charge as a 2-Vector}

In this section we derive a number of consequences of applying the 2-space mathematical machinery to fundamental fermions. Many of the results presented here were arrived at in earlier works by a somewhat different route. In particular, in [7] we began by generalizing the scalar \emph{fermion number} $f$ to a $2\times 2$ real, generally non-Hermitian \emph{matrix} $\bF$ (i.e., $f\to \bF$), and in [8] we arrived at this same matrix $\bF$ by analytically continuing the fermion number operator 
$\Fop\to \bF$.

In the present paper, by contrast, we begin by generalizing \emph{electric charge} (call it $e$) from a \emph{scalar} to a 2-\emph{vector}
$\bQ$
(i.e., $e\to \bQ$) or ``charge vector.'' Only later do we arrive at the matrix $\bF$ described above. The interested reader is encouraged to consult the indicated references for further details regarding these earlier works.

\subsection{The 2-space metric}

If the 2-vector $\bQ=\{q_1, q_2\}$ is assigned to \emph{each} of the matter states $|u\rangle$ and $|d\rangle$, then there must exist a vector (call it $\bQ^c$) that is assigned to \emph{each} of the corresponding anti-matter states $|\overline u\rangle$ and $|\overline d\rangle$, respectively.

Recalling the definition of the charge vector $\bQ$ given in Sec.\ 1.2 (see Eq.\ 8), the (anti) charge vector $\bQ^c$ must be formed in some way from the ordered pair of real numbers $-q_1$ and $-q_2$ corresponding, respectively,
to the electric charges of the $|\overline u\rangle$ and $|\overline d\rangle$ antimatter states. Now, assuming that the scalar product of $\bQ$ and $\bQ^c$, namely, $\bQ\cdot \bQ^c$ should vanish ($\bQ$ and $\bQ^c$ should be orthogonal) so as to insure that matter and antimatter states can be \emph{distinguished} in the 2-space, it follows that no matter what the metric is, as long as it is \emph{real} and \emph{flat}, $\bQ$ and $\bQ^c$ must also be \emph{linearly independent}. To see this, notice first that if $\bQ$ and $\bQ^c$ are \emph{not} linearly independent, then they are, by definition, necessarily, \emph{linearly dependent}, in which case $\bQ^c=-\bQ=\{-q_1, -q_2\}$.

Assuming that the 2-space metric $\boldg$ is \emph{real} and \emph{flat}, $\boldg$ can be represented by the $2\times 2$ matrix
$\boldg  = (g_{11}=1, g_{22}=s, g_{12}=g_{21}=0)$ or
\begin{equation}\label{eqn11}
\boldg = \left( \begin{array}{cc}
1 & 0 \\
0 & s \end{array}\right), 
\end{equation} 
where $|s|=1$.
Then, given $\A\cdot\B=\mathop{\sum}\limits^{2}_{i,j=1}g_{ij}a_ib_j$, where $\A=(a_1,a_2)$ 
is a row-vector and $\B=\{b_1,b_2\}$ is a (conformable) column-vector, the scalar product of $\bQ$ and $\bQ^c=-\bQ$ (Note that $q_1$ is never equal to $q_2$ because these charges correspond to \emph{different} eigenstates of the electric charge operator $\biq$; see Eqs.\ 5--7) is, from (8) and (9), 
\begin{equation}\label{eqn12}
\bQ\cdot \bQ^c = (q_1, q_2)\left(\begin{array}{c}
-q_1 \\
-q_2\end{array}\right) = -q^2_1-sq^2_2,
\end{equation}
which is generally \emph{nonzero} for any $|s|=1$, i.e., $s=\pm 1$ (also see Footnote 9). Therefore,  if $\bQ$ and $\bQ^c$ are to be \emph{orthogonal} $(\bQ\cdot \bQ^c=0)$, the fact that (10) is 
generally nonzero ensures that $\bQ$ and $\bQ^c$ must be \emph{linearly independent}. In this case, given $\bQ=\{q_1, q_2\}$, it must be true that
\begin{equation}\label{eqn13}
\bQ^c = \{-q_2, -q_1\},
\end{equation}
which is \emph{not} proportional to $\bQ$, as required to ensure \emph{linear independence}.

Finally, given the general form for the metric (9), and the linear independence (and orthogonality) of $\bQ$ and $\bQ^c$, one has the \hbox{following result for $s$}
\begin{equation}\label{eqn14}
\bQ\cdot\bQ^c=(q_1, q_2) \left(\begin{array}{c}
-q_2 \\
-q_1 \end{array}\right) = -q_1q_2 - sq_1q_2 = 0,
\end{equation}
if, and only if, $s=-1$. Therefore, the 2-space \emph{metric} is given by
\begin{equation}\label{eqn15}
\boldg = \left( \begin{array}{cc}
1 & 0 \\
0 & -1 \end{array}\right),
\end{equation}
and we see that the requisite 2-space is, necessarily, ``Lorentzian'' or \emph{non-Euclidean} [10].

\subsubsection{Scalar products of 2-vectors}

Using the metric given in (13), and the general formula for $\A\cdot\B$ given in Sec.\ 2.1,  we immediately have the \emph{scalar product} of two different real 2-vectors
\begin{equation}\label{eqn16}
(a, b)\{e,f\}=ae-bf.
\end{equation}
Similarly, the \emph{square} of a real 2-vector is given by 
\begin{equation}\label{eqn17}
(a,b)\{a,b\}=a^2-b^2.
\end{equation}
Here we remind the reader that $(\quad,\quad)$ is a \emph{row} vector while $\{\quad,\quad\}$ is a (conformable) \emph{column} vector. Clearly, the scalar products in (\ref{eqn16}) and (\ref{eqn17}) transform like charge-conjugation-reversing or $\C$-reversing (2-scalar) \emph{charges}.

For example, using (8), (11) and (\ref{eqn17}) we immediately have the \emph{square} of the 2-vectors $\bQ$ and $\bQ^c$, namely,
\begin{equation}\label{eqn18}
\bQ^2=\bQ\cdot \bQ=q^2_1-q^2_2
\end{equation}
and 
\begin{equation}\label{eqn19}
(\bQ^c)^2 = \bQ^c\cdot \bQ^c=q^2_2-q^2_1.
\end{equation}
Therefore,
\begin{equation}\label{eqn20}
\bQ^2= -(\bQ^c)^2,
\end{equation}
which means that $\bQ^2$ and $(\bQ^c)^2$ each transform like $\C$-reversing 2-scalar \emph{charges} [9]. 

It happens that these particular charges can be identified with the \emph{baryon}- or \emph{lepton}-\emph{number} carried by quarks or leptons, respectively (see Ref.\ 7, p.\ 72).
We
will see in a later section that when charge vectors such as $\bQ$ (or $\bQ^c$) are \emph{resolved} in the 
2-space into pairs of linearly independent vectors (e.g., $\bQ=\bU+\bV$), not only are the \emph{components} of $\bQ$, $\bU$ and $\bV$,
$\C$-reversing \emph{charges}, but also given
\begin{equation}\label{eqn21}
\bQ^2=\bU^2+2\bU\cdot \bV+\bV^2,
\end{equation}
$\bU^2$, $2\bU\cdot \bV$ and $\bV^2$ are, like $\bQ^2$, $\C$-reversing charges. The foregoing
collection of 2-scalar charges will be used to define and describe \emph{flavor eigenstates}, \emph{flavor doublets}, and eventually \emph{families} of fundamental fermions.

\subsubsection{A conjectured ``duality''}

Given the number of flavors of quarks and leptons, and an appropriate (renormalizable) Lagrangian, the so-called ``accidental symmetries'' of the Lagrangian [11] are known to ``explain'' the separate conservation of various (global) flavor-defining (Lorentz 4-scalar) ``charges'' [e.g., lepton number, baryon number, strangeness, charm, the third-component of (strong or global) isospin, truth, beauty, electron-, muon-, and tau-numbers]. Now, as demonstrated in detail in 
Ref.\ 7, pp.\ 67--71, given certain real $\C$-reversing scalars---components of various vectors and matrices defined on the internal non-Euclidean 2-space, and various scalar products of 2-%
vectors---it is  possible to describe the flavor eigenstates of fundamental fermions [12].

In principle, what one does is to identify the mutually-commuting $\C$-reversing
2-space ``charges'' (call them $C_i$) or charge-like quantum numbers associated with a particular flavor, and then write the corresponding simultaneous flavor-eigenstate as
 \begin{equation}\label{eqn22}
|C_1, C_2, C_3, \ldots, C_n\rangle.
\end{equation}
Here $C_1, C_2, C_3, \ldots, C_n$, are said to be the ``good'' charge-like quantum numbers (charges) associated with a particular flavor [13]. 
Now it also happens that these \emph{observable} real numbers can be identified with quantum numbers such as \emph{electric charge, lepton number, baryon number, strangeness, charm}, the 
third-component of (strong or global) \emph{isospin}, \emph{truth} and \emph{beauty} (see
Ref.\ 7, p.\ 72). In short, \emph{these} 2-space \emph{charges look very much like those associated with the aforementioned ``accidental symmetries'' of the Lagrangian}!
The foregoing properties of the non-Euclidean charge-like scalars, leads naturally to the following ``duality'' conjecture:
\medskip

\emph{The global} (\emph{flavor-defining}) \emph{charges associated with the ``accidental symmetries'' of the Lagrangian describing strong and electroweak interactions, and the global} (\emph{flavor-defining}) \emph{charges associated with the non-Euclidean 2-space, are} (\emph{essentially}) \emph{one and the same charges.}

\subsection{Charge conjugation in the 2-space}

Given that there are numerous $\C$-reversing scalars in the 2-space (see Sec.\ 2.1.1), there must exist a $2\times 2$ matrix, call it $\X$, that serves to transform these \emph{scalars}, various 2-\emph{vectors} such as $\bQ$ or $\bQ^c$, and various $2\times 2$ \emph{matrices}, to their corresponding $\C$-reversed (2-space) counterparts. In particular, a matrix $\X$ should exist such that (use Eqs.\ 8 and 11)
\begin{equation}\label{eqn23}
\X\bQ=\bQ^c
\end{equation}
and
\begin{equation}\label{eqn24}
\X\bQ^c=\bQ.
\end{equation}
From (21) and (22) it follows that $\X$ must equal its multiplicative inverse $(\X=\X^{-1})$, and thus
\begin{equation}\label{eqn25}
\X^2=\I_2,
\end{equation}
where $\I_2$ is the $2\times 2$ identity matrix.

Write $\X$ in the general form ($\X$ is real)
\begin{equation}\label{eqn26}
\X = \left(\begin{array}{cc}
a & b \\
c & d \end{array}\right),
\end{equation}
and consider the situation where one of the two charges (say $q_2$) associated with $\bQ=\{q_1, q_2\}$ is \emph{zero}, and the other charge $(q_1)$ is \emph{nonzero} (this is actually the case for leptons).

In this particular case, we have (use Eqs.\ 8, 11, 21 and 24)
\begin{equation}\label{eqn27}
\left(\begin{array}{cc}
a & b \\
c & d \end{array}\right)\left(\begin{array}{c}
q_1 \\
0 \end{array}\right) = \left(\begin{array}{c}
0 \\
-q_1 \end{array}\right),
\end{equation}
which means that
\begin{equation}\label{eqn28}
aq_1=0
\end{equation}
and
\begin{equation}\label{eqn29}
cq_1=-q_1.
\end{equation}

And therefore, since $q_1\ne 0$, it must be true from (26) and (27) that $a=0$ and $c=-1$.

 Since 
$\X\bQ^c=\bQ$ it must also be true that
\begin{equation}\label{eqn30}
\left(\begin{array}{cc}
0 & b \\
-1 & d \end{array}\right)\left(\begin{array}{c}
0 \\
-q_1 \end{array}\right) = \left(\begin{array}{c}
q_1 \\
0 \end{array}\right),
\end{equation}
which means that
\begin{equation}\label{eqn31}
-bq_1 = q_1 
\end{equation}
and
\begin{equation}\label{eqn32}
-dq_1 = 0.
\end{equation}
Finally, since $q_1\ne 0$ it must be true from (29) and (30) that $b=-1$ and $d=0$.

Collecting the foregoing matrix elements, we have
\begin{equation}\label{eqn33}
\X = \left(\begin{array}{cc}
0 & -1 \\
-1 & 0 \end{array}\right),
\end{equation}
or
\begin{equation}\label{eqn34}
\X=-\bfsigma_\X,
\end{equation}
where $\bfsigma_\X$ or $\bfsigma_1$ is one of the familiar Pauli matrices. In general, the matrix $\X=-\bfsigma_\X$ should apply (in the 2-space) to 2-\emph{scalars}, 2-\emph{vectors},
$2\times 2$ \emph{matrices}, and to both \emph{quarks} and \emph{leptons}.

\subsubsection{Transformation of the metric and other matrices under $\X$}

Any $2\times 2$ matrix $\M$, appropriate to the 2-space description of fundamental fermions, should transform under $\X=-\bfsigma_\X$ to its $\C$-reversed counterpart $\M^c$ according to the \emph{similarity transformation}
\begin{equation}\label{eqn35}
\X\;\M\;\X^{-1}=\M^c,
\end{equation}
or because $\X=\X^{-1}=-\bfsigma_\X$, equivalently as
\begin{equation}\label{eqn36}
(-\bfsigma_\X)\M(-\bfsigma_\X)=\M^c.
\end{equation}
For example, the metric $\boldg$ (see Eq. \ref{eqn15}) is found to be $\C$-\emph{reversing} since
\begin{equation}\label{eqn37}
(-\bfsigma_\X)\boldg(-\bfsigma_\X)=-\boldg.
\end{equation}

A matrix that is $\C$-\emph{invariant} (e.g., the matrix $\X$) would, necessarily, have the form
\begin{equation}\label{eqn38}
\N=\left(\begin{array}{cc}
a& b \\
b & a \end{array}\right),
\end{equation}
where it is clear that
\begin{equation}\label{eqn39}
(-\bfsigma_\X)\N(-\bfsigma_\X)=\N.
\end{equation}

Now let us apply the foregoing similarity transformation to the matrix $\F$, which represents the \emph{generalized fermion number} in this 2-space [see Ref.\ 7, pp.\ 4--12].

\subsection{The generalized fermion number $\F(v)$}

There are other $2\times 2$ matrices besides $\X$ that act on the 2-vectors $\bQ$ and $\bQ^c$. Let us \emph{define a matrix} $\F$ \emph{whose eigenvalues} ${\mathrm{f}}$ \emph{are the fermion numbers} ${\mathrm{f}}_m=+1$ \emph{for matter, and} ${\mathrm{f}}_a=-1$ \emph{for antimatter, and whose eigenvectors are  the electric-charge vectors}
$\bQ$ \emph{and} $\bQ^c$, \emph{respectively}. Then
\begin{equation}\label{eqn40}
\F\bQ=f_m\bQ
\end{equation}
and
\begin{equation}\label{eqn41}
\F\bQ^c = f_a\bQ^c.
\end{equation}
Clearly, (38) and (39) require $\F$ to be equal to its multiplicative inverse, i.e., $\F=\F^{-1}$ or $\F^2=\I_2$.

Given that the eigenvalues of $\F$ are $f_m$ and $f_a$, the diagonal form for $\F$ is simply 
(see Ref.\ 7, p.\ 4)
\begin{equation}\label{eqn42}
\F_{\diag} = \left( \begin{array}{cc}
f_m & 0 \\
0 & f_a \end{array}\right).
\end{equation}
And, from (31) and (33) the $\C$-reversed counterpart of $\F_{\diag}$ is, necessarily, given by the \emph{similarity transformation}
\begin{equation}\label{eqn43}
(-\bfsigma_\X)\F_{\diag}(-\bfsigma_\X)=-\F_{\diag}.
\end{equation}
From the minus sign on the right hand side of (41) we see that the scalar fermion numbers $f_m$ and $f_a$ properly change signs (are $\C$-reversing ``charges'')
under $-\bfsigma_\X$. 

Consider next a more general (nondiagonal) matrix $\F$.
Because the \emph{trace, determinant} and \emph{square} of $\F$ are \emph{invariants}, one has
\begin{align}
tr \F & = tr \F_{\diag} = f_m+f_a=0 \label{eqn44} \\ 
det \F & = det \F_{\diag} = f_m\cdot f_a = -1 \label{eqn45}\\ 
\intertext{and}
\F^2 & = \F^2_{\diag}=\I_2. \label{eqn46}
\end{align}
And, because $\F$ is traceless it must have the general form
\begin{equation}\label{eqn47}
\F = \left( \begin{array}{cc}
a & b \\
c & -a\end{array}\right).
\end{equation}

Finally, because $\F$ should transform in the same way as $\F_{\diag}$ under $\X=-\bfsigma_\X$ we have the similarity transform
\begin{equation}\label{eqn48}
(-\bfsigma_\X)\F(-\bfsigma_\X)=-\F.
\end{equation}
Using $\F$ as expressed by (45) and employing (46), one finds that $c=-b$. Hence, the most \emph{general} form of $\F$ is, necessarily,
\begin{equation}\label{eqn49}
\F = \left(\begin{array}{cc}
a&b \\
-b&-a \end{array}\right),
\end{equation}
where $a^2-b^2=1$ since $det \F=-1$.

Making the following substitutions in (47), namely,
\begin{equation}\label{eqn50}
a=\cosh v
\end{equation}
and
\begin{equation}\label{eqn51}
b=\pm\sinh v,
\end{equation}
where $-\infty\le v\le +\infty$ is a \emph{real} (dimensionless) 
parameter in the range indicated, $\F$ or $\F(v)$ finally assumes the general form
\begin{equation}\label{eqn52}
\F(v) = \left( \begin{array}{rl}
\cosh v, & \pm \sinh v \\
\mp\sinh v, & -\cosh v\end{array}\right).
\end{equation}
Here, $\F^2(v)=\I_2$ for any $v$, $\F(v)$ satisfies the boundary condition $\F(0)=\F_{\diag}$, and the non-Euclidean 2-scalars $\bQ^2$ and $(\bQ^c)^2$ are left 
\emph{invariant} by the transformation $\F(v)$.

\subsection{Distinguishing quarks and leptons}

Choosing the upper signs in (\ref{eqn52}), the matrix $\Fv$ becomes
\begin{equation}\label{eqn53}
\F(v) = \left( \begin{array}{cc}
\cosh v & \sinh v\\
-\sinh v & -\cosh v
\end{array}\right),
\end{equation}
where $v$ is also chosen to be a positive real number (see Ref.\ 7, p.\ 50 and 54).

As described in Ref.\ 7, pp.\ 52--55, 
the parameter $v$ distinguishes between \emph{quarks} and
\emph{leptons}.  In particular, the 
parameter $v$ is found to be \emph{quantized} and obeys the ``quantum condition":  
\begin{equation}\label{eqn54}
v=\ln M_c,
\end{equation}
where $M_c$ counts both the \emph{number} of fundamental fermions in a
strongly-bound composite 
fermion, and the \emph{strong-color multiplicity}.  That is, $M_c = 3$ for quarks
(strong-color triplets) and 
$M_c = 1$ for leptons (strong-color singlets).  Thus we have found that a connection exists between the 2-space description of quarks and leptons, and their associated 
\emph{strong} colors!

\subsubsection{Quark and lepton electric charges and $B$ and $L$}

It has been shown (see Sec.\ 2.3, Eqs. \ref{eqn40} and \ref{eqn41}; Ref.\ 7, pp.\ 52--55,
and Ref.\ 8) that the quark and lepton electric charges are the 
``up''-``down'' components of the eigenvectors of the matrix $\Fv$ specified by (51) and (52). In particular, the quark charges are given by $(M_c=3)$
\begin{eqnarray}
q_1(f) & = & \frac{(M^2_c-1)}{2M_c(M_c-f)} = +\frac{2}{3}\hbox{ for }f=+1\hbox{ and }+\frac{1}{3}\hbox{ for }f=-1, \label{eqn55} \\
q_2(f) & = & q_1(f)-1, \label{eqn56}
\end{eqnarray}
where the baryon number for quarks is $B=q^2_1(f)-q^2_2(f)=\pm\frac{1}{3}$ for $f=\pm 1$.
Similarly, the lepton electric charges are given by $(M_c=1$)
\begin{eqnarray}
q'_1(f) & = & \frac{-(M_c^2-1)}{2M_c(M_c-f)} = -1\hbox{ for }f=+1\hbox{ and }0\hbox{ for }f=-1, \label{eqn57} \\
q'_2(f) & = & q'_1(f)+1, \label{58}
\end{eqnarray}
where the lepton number for leptons is $L=[q'_1(f)]^2-[q'_2(f)]^2=\pm 1$ for $f=\pm 1$.

In summary, the new 2-space description, and 
$\Fv$, is found to provide an 
explanation for the quark-lepton ``dichotomy" of fundamental fermions in
addition to the matter-antimatter, and ``up"-`` down" type flavor-dichotomies.  

\subsection{Representing flavor doublets in the 2-space}

Consider again the eigenvectors $\bQ$ of $\Fv$ for fundamental fermions.
Since the space on which $\Fv$
``acts" is two-dimensional, an \emph{observable} vector $\Q$ can
be ``resolved" into two (no more or less)
\emph{observable},  linearly-independent vectors, call them $\U$ and $\V$,
as
$\Q = \U + \V$ [14].  Now, because these
three vectors ($\Q$, $\U$, and $\V$) are \emph{simultaneous}
observables, it makes sense to speak of this ``triad"
of vectors as being a well defined geometric object, namely, a ``vector triad."

Recognizing that the components of $\Q$, $\U$ and $\V$ are $\C$-reversing
charge-like \emph{observables} [13]
 we can
write these \emph{observable} ``charge" vectors as
\begin{eqnarray}
\Q & = & \{q_1, q_2\}\label{eqn59} \\
\U & = & \{u_1, u_2\}\label{eqn60} \\
\V & = & \{v_1, v_2\},\label{eqn61}
\end{eqnarray}
where $q_1$, $q_2$, $u_1$, $u_2$, $v_1$ and $v_2$ are the various
\emph{observable} ``charges" (e.g., $q_1$ and $q_2$ are the \emph{electric} charges). 
Given $\Q = \U + \V$, the non-Euclidean metric (\ref{eqn15}), and Eqs. (\ref{eqn59}) through
(\ref{eqn61}), we
 find the 
associated \emph{observable} quadratic-``charges" [13]
\begin{eqnarray}
\Q^2 & = & \U^2 + 2 \U\bullet\V + \V^2\label{eqn62} \\
2\U\bullet\V & = & 2(u_1v_1-u_2v_2)\label{eqn63} \\
\U^2 & = & u^2_1 - u^2_2 \label{eqn64}\\
\V^2 & = & v^2_1-v^2_2.\label{eqn65}
\end{eqnarray}
Finally, using the foregoing collection of (global) flavor-defining charges, we can express the \emph{two} quantum
states (simultaneous 
flavor-eigenstates) associated with a \emph{single} vector-triad in the
form of ``ket" vectors as follows (Ref.\ 7, pp.\ 16--18)
\begin{equation}\label{eqn66}
\left.
\begin{array}{l}
|q_1, u_1, v_1, \Q^2, \U^2, 2\U\bullet \V, \V^2\rangle \\[2mm]
|q_2, u_2, v_2, \Q^2, \U^2, 2\U\bullet \V, \V^2\rangle
\end{array}\right\}\lower.15in\hbox{.}
\end{equation}
Here, the state $|q_1, u_1, v_1, \Q^2, \U^2, 2\U\bullet \V, \V^2\rangle$
 represents the ``up"-type flavor-eigenstate, and
$|q_2, u_2, v_2, \Q^2, \U^2, 2\U\bullet\V, \V^2\rangle$
 represents the corresponding ``down"- type flavor-eigenstate
in a flavor doublet of fundamental fermions [12, 15].

\subsection{Family replication and the number of families}

In Ref.\ 7, pp.\ 39--49 and pp.\ 59--65, it is shown that flavor doublets (hence families) are \emph{replicated} and that there are only \emph{three} families of quarks and leptons. We  refer the reader to [7] for a full and detailed account. Here we simply outline how this situation comes about.

Once again, by the definition of a linear-vector 2-space, a 2-vector such as $\Q$ can always be resolved into a pair (no more, or less) of linearly-independent vectors $\U$ and $\V$ as $\Q=\U+\V$ (see Sec.\ 2.5). And, since $\Q$ \emph{represents} a flavor doublet, so should $\U$ and $\V$ \emph{represent} this \emph{same} flavor doublet. But, if this is so, \emph{different vector-resolutions of $\Q$ 
{\rm(}i.e., different vector-triads{\rm)} should correspond to different flavor-doublets having the same $\Q$}. In other words, flavor doublets should be \emph{replicated}.

Since $\Q$ can be resolved (mathematically) in an infinite number of ways, we might suppose that there are an \emph{infinite} number of flavor doublets, and hence, families. But, because of various ``quantum constraints,'' it is possible to show that $\Q$ \emph{can be resolved in only three physically acceptable ways for $\Q$-vectors associated with either quarks or leptons}. In other words, there can be only six quark flavors and six lepton flavors, which leads to the 
(\emph{ex post facto})``prediction'' of three quark-lepton families.

\subsection{New internal differences and neutrino mixing}

It is important to understand that the new 2-space description of fundamental fermions (quarks and leptons) provides a distinction between these particles that goes beyond differences that can be explained by mass differences alone. For example, in the standard model the only difference between the $u$, $c$ and $t$ quarks is that they have different masses. Otherwise, these particles experience identical strong and electroweak interactions. Moreover, as described in Section 2.1.2, the separate conservation of quantum numbers such as ``charm'' and ``truth'' can be attributed to certain unavoidable ``accidental symmetries'' associated with the (renormalizable) Lagrangian describing the (strong) interactions of these particles [11].

Taken at face value, these accidental symmetries would seem to imply that there are no internal ``wheels and gears'' that would distinguish a $u$ quark from a $c$ quark, for example. But, if the string theories are correct, these particles would be associated with different ``handles'' on the compactified space (see Ref.\
16, Vol.\ 2, p.\ 408), and so would be different in this \emph{additional} sense. Likewise, in the present non-Euclidean
2-space description, topological differences in addition to a variety of (global) 2-scalars, which are only \emph{indirectly} related to the  accidental symmetries of the Lagrangian, serve to provide further distinctions between matter particles.

A possible experimental signal of such ``internal'' differences is to be found in the recent observations at the Super Kamiokande of bi-maximal neutrino mixing [17]. Models which begin by positing a neutrino mass-matrix and associated mixing-parameters, such as the three-generation model proposed by Georgi and Glashow
[18], do an acceptable job of describing the observations. However, bi-maximal mixing may have a deeper explanation in terms of internal topological differences (in the non-Euclidean 2-space) between $\nu_e$, and $\nu_\mu$ or $\nu_\tau$ neutrinos.

With respect to the internal transformation $\Fv$, the topology of the non-Euclidean ``vector triad'' (see Sec.\ 2.6, Ref.\ 7, p.\ 57, Ref.\ 20, and the qualifying remarks in Footnote 19) representing the $\nu_e$ ($\nu_\mu$ or $\nu_\tau$), is found to be that of a cylinder (M\"obius strip). And, assuming that a change in topology during neutrino mixing is suppressed by energy ``barriers,'' or other topological ``barriers'' (e.g., one cannot continuously deform a doughnut into a sphere), while neutrino mixing without topology-change is (relatively) enhanced, one can readily explain the experimental observation of (nearly) maximal $\nu_\mu-\nu_\tau$ neutrino mixing---at least maximal $\nu_\mu-\nu_\tau$ mixing over long distances, where the proposed topological influences are expected to be cumulative [19, 20]. If this qualitative explanation is basically correct, then it follows that the neutrino mass-matrix and associated 
mixing-parameters needed to explain bi-maximal neutrino mixing, would be the \emph{result}, at least in part, of these deeper (internal) topological differences between neutrinos, and not their \emph{cause}.

\section{Summary and Conclusions}

By insisting that the electric charge associated with $SU(2)$ flavor doublets, (i.e., weak isospin doublets) of fundamental fermions (quarks and leptons) can be treated as a \emph{real, internal} 2-\emph{vector}, 
considerations of self consistency dictate that all of the mathematical ``machinery'' associated with the linear vector 2-space (e.g., the metric, and various transformations) must be brought to bear when describing fundamental fermions. While this mathematical machinery is very simple, its application to fundamental fermions immediately leads to a (modest) extension of the standard model description of quarks and leptons, and to a number of predictions. 

The model predicts, among other things, that unlike the situation in the standard model [3, 6], particles such as the $u$, $c$ and $t$ quarks are characterized by significant \emph{internal} (topological and other) differences. Similar differences may help explain recent observations of (nearly) maximal $\nu_\mu-\nu_\tau$ mixing [17--20]. Moreover, while we began this paper with the introduction of an $SU(2)$ flavor doublet of ``quarks'' $|u\rangle, |d\rangle$ and their antiparticle counterparts (see Sec.\ 1.0), very little else was assumed about \emph{quarks}, \emph{leptons}, \emph{internal} (\emph{strong}) 
\emph{colors} or \emph{family replication}. And yet,  the subsequent 2-space description
correctly predicts that all of these things, and more, are properties of fundamental fermions.
For example, to maintain compatibility with the standard model, we were forced to (tentatively) conclude that certain global ``charges'' associated with the ``accidental symmetries'' of the Lagrangian describing strong and electroweak interactions, and various 2-space charges, are (essentially) one and the same charges (see the ``duality'' conjecture in Sec.\ 2.1.2). Finally, family replication emerged here as little more than the number of different physically acceptable ways (\emph{three} physically acceptable ways are predicted) the vector $\bQ$ for quarks and leptons can be \emph{resolved} in the 2-space (see Sec.\ 2.6 and Ref.\ 7, pp.\ 39--49) into
pairs
of
linearly
independent ``basis'' vectors $\bU$ and $\bV$ (i.e., $\Q=\bU+\bV$).

In closing it should be pointed out that popular extensions of the standard model such as the so-called realistic (free fermionic) three-generation string models [21, 22], also provide an ``explanation'' for \emph{family replication} and the \emph{number} of families. Does this mean that in spite of very significant and obvious differences there are, nevertheless, ``deep'' connections between the proposed 2-space description and string theories?

I thank R.\ Zannelli for pointing out a problem with the existing 2-space description of muons (see Tables II and IV, and p.\ 57 in Ref.\ 7). This problem, together with my proposed ``remedy,'' are briefly described in [19].

\renewcommand{\refname}{References and Footnotes}


\begin{thebibliography}{99}
\bibitem[9]{9}Note that if $\bQ^c=-\bQ$, then $(\bQ^c)^2=(-\bQ)^2$ no matter what the
metric is (i.e., $s=\pm 1$). That is, if $\bQ^c=-\bQ$, $(\bQ^c)^2$ and
$\bQ^2$ are the \emph{same} for matter and antimatter, so they do \emph{not}
transform like \emph{charges}, and consequently do \emph{not} distinguish
between matter and antimatter.
\bibitem[12]{12}  When weak interactions are ``turned off'' flavor eigenstates and mass 
eigenstates are one and the same.  For the most part, when we speak here of flavor eigenstates, we are referring to the situation where flavor eigenstates
and mass eigenstates are the same.
\bibitem[14]{14} When we say that the vectors $\bQ$, $\bU$ and $\bV$ are \emph{observables}, we mean that their associated component-``charges'' are mutually-commuting simultaneous observables. Hence, all of these charge-like components can be known in principle, at the \emph{same} time, meaning that the vectors $\bQ$, $\bU$ and $\bV$ can be known simultaneously.
Thus the vector ``triad'' $(\Q, \bU, \bV)$ is a well defined geometric object.
\end{thebibliography}
\end{document}